# GENERATING PYTHON CODE FROM OBJECT-Z SPECIFICATIONS

A.F. Al Azzawi[1], M. Bettaz[2] and H. M. Al-Refai[2]

[1]Department of Software Engineering, IT Faculty, Philadelphia University, Amman, Jordan
[2]Department of Computer Science, IT Faculty, Philadelphia University, Amman, Jordan

## ABSTRACT

*__Object-Z__ is an object-oriented specification language which extends the Z language with classes, objects, inheritance and polymorphism that can be used to represent the specification of a complex system as collections of objects. There are a number of existing works that mapped Object-Z to C++ and Java programming languages. Since Python and Object-Z share many similarities, both are object-oriented paradigm, support set theory and predicate calculus moreover, Python is a functional programming language which is naturally closer to formal specifications, we propose a mapping from Object-Z specifications to Python code that covers some Object-Z constructs and express its specifications in Python to validate these specifications. The validations are used in the mapping covered preconditions, post-conditions, and invariants that are built using lambda function and Python's decorator. This work has found Python is an excellent language for developing libraries to map Object-Z specifications to Python.*

## KEYWORDS

*Object-Z, Python, Object Oriented Programming, Formal Language Specification, Design by Contract*

## 1. INTRODUCTION

The formal languages Object-Z, VDM++, and UML-B are based on object-oriented styles which often mentioned in the literature. According to the comparison has been given in [1], the Object-Z has an important feature over other languages, which are:

−Powerful semantic and calculus (predicate calculus and set theory)
−strong support of objects
−its specification style corresponds directly to object oriented programming constructs while UML-B weakly supports the concepts "object" and VDM++ does not have exact formal calculus.

However, performing proofs through formal languages is a difficult task and requires a lot of skills in mathematics and an alternative approach is to map Object-Z to programming language and validates the specifications during execution. There are many works that mapped Object-Z specifications to object oriented programming like C++ and Java, some presents a mapping method to cover some Object-Z constructs like basic type definition, schema, class, state schema, operation operators, state schema predicates. Types of constants, methods and template class [2], [3], [4], [5] and other works tried to complete these constructs such as all types of definitions, object aggregation, object containment and other operation operators [1].

Also, there are works that mapped Object-Z specifications to object-oriented specification language which incorporating specification concepts and such linkage would allow system requirements to be specified in a high-level formal language, but validated at program language level. Like Spec# specification language which is an extension to .Net C# language [6], Perfect Developer which is an OO language that supports verification and validation [7], and MathLang which is a system for computerizing mathematical texts to check the correctness. MathLang is used to test the computerization of formal specifications written in Object-Z [8].

unlike C++ or Java programming, which is used for Object-Z mapping, Python supports multiple paradigms, dynamic typing, high-level built-in data types, and many add-on packages available that extends its capabilities with numerical computations, and scientific graphics.(SciPy,nNumPy) [9, 10] and also it incorporates an predicate calculus, set theory, and theorem proven to validate the written specifications with it. This makes it suitable for scientific and formal specifications mapping and this work mapped a part of Object-Z specification into program level Python specifications.

## 2. BACKGROUND

In this section, we review main constructs of Object-Z, why we choose Python programming to map Object-Z specifications and then, considering Design by Contract methodology to validate contracts in mapping Object-Z to Python.

### 2.1. OBJECT-Z

Object-Z is an extension of Z language by the adding of an object-oriented paradigm constructs such as classes and other object-oriented notions such as polymorphism and single/multiple inheritance. While Z is based on mathematical notation such as set theory, lambda calculus, and first order predicate and Z's expressions are typed which is used mathematical functions and predicates.

In Object-Z, class definition comprises a named box with optional formal general parameters that introduce a basic type only used in an expression or predicate, and also may have a visibility list, inherited classes, local type and constant definitions, at most one state schema, associated initial state schema and operations[12], [13], [14] and the following its basic class structure:

ClassName[general parameters]
    constant definitions
    type definitions
    state schema
    initial state
    schema operations

The following figure is an example of Object-Z specification for simple credit card bank account system described in [13], Each account has two numbers (current balance and credit limit) and three operations (withdrawn, deposited, and withdrawavil). This Object-Z specification example will be transformed step by step to Python code in next sections.

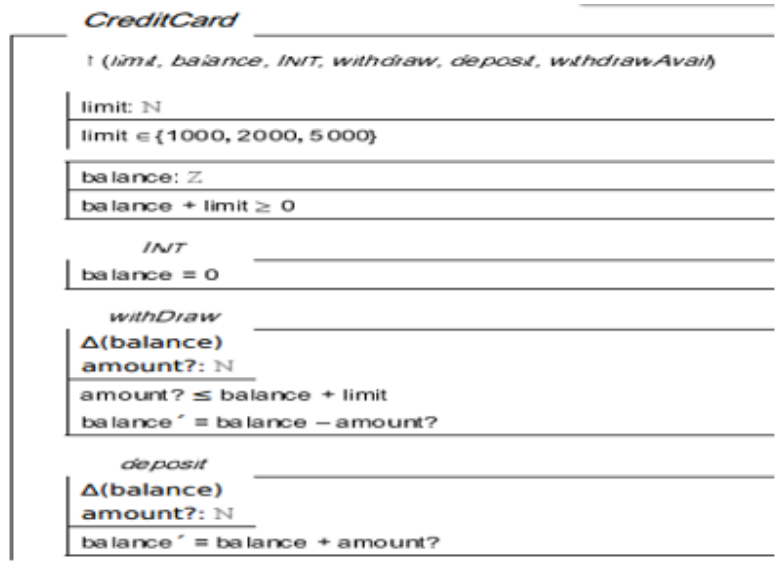

Figure 1. Object-Z credit card specification example

## 2.2. PYTHON

Python is a powerful programming language that supports multiple paradigms, dynamic typing, automatic memory management, high-level built-in data types, full modularity, hierarchical packages and many add-on packages available that extends its capabilities with numerical computations, scientific graphics, and graphical user interface programming. This makes it very useful for scientific programming [9, 10].

Python has an extraordinarily simple syntax and reading Python program is like reading English, and program written in Python is only half as long as written in C, C++, or Java [15]. Python allows concentrating on the solution of the problem rather than the language itself [15], and due to its open-source nature, Python code can be ported to many platforms without requiring any changes at all if you avoid any system-dependent features.

Python and Object-Z language have many similarities. Both of them are based on object-oriented paradigms, incorporates predicate calculus, and set theory moreover, Python is a functional programming language which is naturally closer to formal specifications.

## 2.3. DESIGN BY CONTRACT

The Design by contract is a software correctness methodology used to control the interaction between modules by precise specifications based on the principle of preconditions and post-conditions to ensure consistence between the program and the specification [11].
Design by Contract plays positive effects on design, and testing [16]. For that, many efforts spent to incorporate this methodology into different object oriented programming languages. For Python programming, there are many designs by contact packages like PyContract, PyContracts which allows annotating functions with contract expressions, but these packages used a syntax which does not allow richer contract expressions, does not express invariants to class attributes, and there are only contracts deal with local scope variables of the function.

In this paper, we use run time validation contracts in mapping Object-Z to Python that will cover preconditions, post-conditions, and consistency constraints (invariants) to check complex conditions using lambda functions. Python provides decorators which can be used to implement similar functionality of these contracts, and these decorators are wrappers around a given class or method.Which describes a formal language that can be used to define model constraints in terms of predicate logic.

Contracts on functions consist of preconditions and post-conditions, In this work, A precondition decorator "pre" is implemented to specify what must be true upon entrance to the function, while post-condition decorator "pos" is implemented to specify what must be true after the function has successfully returned, and these "pre" and "pos" decorators can be used on class methods too. Moreover, the invariant decorator "inv" is implemented to specify a condition always true for class instances before any method call and after its return. "pre", "pos" and "inv" decorators take a lambda function as argument to validate constraints upon class or method.

## 3. MAPPING OBJECT-Z TO PYTHON

This section provides mapping rules to transform basic Object-Z contracts to Python contracts and shows how to implement these transformations through Object-Z credit card specification example given in Figure 1.

### 3.1. CLASS DEFINITION

Class schema in Object-Z can be refereed to notion of a class by encapsulating local type and constant declarations, initial state schema, and zero or more operation schemas for the given state[12], [13].

**Rule1:** Every Object-Z class schema is equivalent to a Python class declaration with same name and no consideration to any formal general parameter types if exits, since Python is a dynamic type checking language that verifies information type safety at run time, and the following shows the mapping for class definition found in figure 1:

```
class CreditCard:
    …
```

### 3.2. VISIBILITY LIST

All members found in Object-Z visibility list class are visible to the environment of the object of all classes, and members which is not found in visibility list are visible only to the object of class and all its derived classes[12]. In Python, all class members will be public by default and otherwise will be private if it's name prefixed with two underscores which is not accessed from outside the class.

**Rule2:** All members of a class not found in the visibility list will be mapped to private by prefixing its name with two underscores which can be accessed in same class through it's name or in derived classes through class parent name, otherwise it is public by default.

### 3.3. CONSTANT DECLARATIONS

In Object-Z, constant declarations are related to a class and have fix values not changed by any operations of the class, but it may be differ in objects of the class [12], [13]. Since writing function is easier than writing class and Python objects can act as an object of function, we map these type declarations to functions which are decorated by type condition validation.

**Rule3:** Each constant declaration will be mapped to a function decorated with "pre" predicate to check the constant type and its initialization**.**

**Example:** Suppose the following constant declaration found in figure 1:

limit: N

The type of natural numbers will be mapped to a function with precondition checks if the constant of type integer and greater than or equal zero, as follows:

@pre(lambda:isinstance(n, int) and n >= 0)
def N(n): return n

The above limit declaration will be mapped to

limit=N

which means that constant "limit" and function "N" will have same reference.

### 3.4. CONSTANT SCHEMA PREDICATE

The constant schema predicate is a predicate  that gives a correct value for constant in  schema [14].

**Rule4 :** The constant schema predicate will be mapped to "inv" decorator which uses lambda function to check  constant correct value.

**Example:** Suppose that we want to map the following constant schema predicate found in the CreditCard class in figure 1:

limit ε {1000, 2000,3000}

This will be mapped as follows:

@inv(lambda:limit in set(1000, 2000,3000)):
class  CreditCard:
        …

### 3.5. STATE SCHEMA

The state schema is a construct that declares attributes correspond to the class variables, and determines correct relationship between their values through predicate [13], [14].

**Rule5:** Each variable found in the state schema will be mapped to class attribute and its state predicate will be considered as an invariant class decoration "inv" .

Primary variables are either visible or invisible, invisible variables (attributes) require to start with two underscores symbol so cannot modified from outside the class. If a variable is associated with invariant to ensure its correctness, the schema must have invariant to check the value of the attribute.

**Example:** Suppose that we want to map the state schema found for the class in figure 1 which has an integer balance and the invariant that specifies balance + limit must be greater than or equal to zero. This state schema will mapped to:

@inv(lambda:balance +limit>=0)
class  CreditCard:
        balance=Z
…

where Z a function with a precondition to ensure if the constant is of type integer and greater than or equal zero as follows:

@pre(lambda:isinstance(n, int) and n >= 0)
def Z(n): return ntion.

### 3.6. INITIAL STATE SCHEMA

The initial state schema has no declaration part and its predicates restrict the possible values of the state variables and constants of the class [12]. The initial schema determines the initial state of all created objects and always named INIT.

**Rule6:** The initial state schema will be mapped to __init__ constructor that specifies the initial attributes values  for all created objects.

**Example:** The initial state schema for Credit Card in figure 1 that sets the balance attribute to zero will be mapped to:

def __init__(self):
    self. balance=0

### 3.7. OPERATION SCHEMA

Operation schema changes the object state from one to another and its declaration starts with a delta list followed by communication inputs/outputs passed between the object and its environment, followed by predicates specifying the constraints on state variables [12].

**Rule7:** Each operation schema will be mapped to a Python method with its associated inputs declaration, and its predicate list will be mapped to set of lambda preconditions and lambda post-conditions.

**Example:** The withdraw operation schema for CreditCard in figure 1 will be mapped to:

```
@pre(lambda: amount<=self.balance+self.limit)
def withdraw(self, amount: N):
    self.balance= self.balance-amount
```

### 3.8. OBJECT INTERACTIONS

Class construct may have individual objects. Figure 2 shows the specification of the class Two Cards that contains two instances of the CreditCard class, deposit, withdraw money from the different cards, and transfer money between them[13].

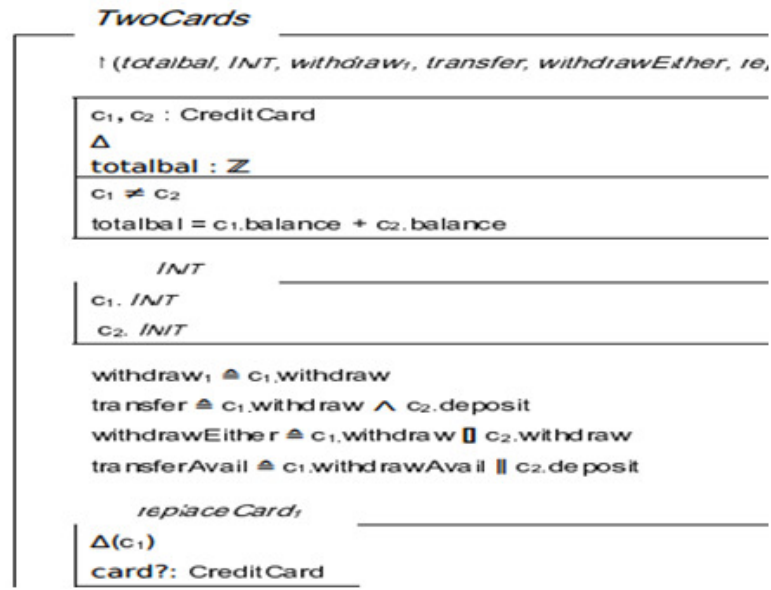

Figure2. Object-Z Two credit cards specification example

### 3.8.1. Secondary variables

Secondary variables may changed in terms of their primary variables found in the class, so that any change in the primary variables will affect the secondary variables and the operation that changed primary variables must update the secondary variables[13].

**Rule8:** Write a function to update the secondary variable in case of any change happened in the value of primary variables and decorate the class with this function. This updated function will be called after calling any method in the class to change the secondary variable according to invariant state that relates these variables.

**Example:** Suppose we want to map the state scheme of Two Cards class in figure2 that has two instances of type CreditCard and the secondary variable total balance defined in term of primary variable balance found the instances of CredirCard.

This state schema will be mapped to:

```
def findtotal(self):
    if elf.c1!=self. c2:
        totalbalance=self.c1.balance+self. c2.balance
```

```
@decorate_all(findtotal)
class  TwoCards:
        totalbalance=Z
…
```

The decorate all function implementation is found in our library which is used to call the function found between braces after calling any method in the class that changes the secondary variables.

### 3.8.3. Initial schema referencing

object.__init__(self) is a call to the initial schema  for of the "object",  the init schema for an object  of  Two Card class shown in figure 2 will be mapped using rule 6 as follows:

```
def __init__(self):
    c1.__init__(self)
    c2.__init__(self)
```

### 3.8.4. Non-deterministic choice

The non-deterministic choice operator is used to model the occurrence of at most one of a pair of operations and when both operations are enabled, the operation will be chosen non-deterministically [13].

OperationExpression ::= OperationExpression [] OperationExpression

**Rule9:** Non-deterministic choice operator in the above form will be mapped to

```
def op1(**kwargs):
        #body
def op2(**kwargs):
        #body

op3 = choice(op1, op2)
```

  where

```
def choice(sh1, sh2, **kwargs):
    r=random.randint(0,1)
    r1=lambda **kwargs:sh1(**kwargs)
    r2=lambda **kwargs:sh2(**kwargs)
    L=[r1,r2] if x==0 else L=[r2,r1]
    return L[0] if L[0] else return L[1]
```

The **kwargs  is  a dictionary for the inputs  of the two  schemes.

each input can be accessed through the dictionary, for example to access the value of  x, we write kwargs[x].

**Example:** The  Operation  Expression  withdrawEither? c1.withdraw  []  c2. withdraw found in scheme of TwoCards class in figure2 will be mapped to:

withdrawEither = choice(c1.withdraw, c2.withdraw)

### 3.8.5. Sequential composition

The sequential composition operator is equivalent to performing the first operation followed by the second and the first operation output variables are identified and equated with the input variables of the second operation having the same base names, which means that sequential composition is equivalent to the parallel composition mapping [13].
OpExp3 ::= OpExp1 ; OpExp2

**Rule10:** Sequential composition operator in the above form will be mapped to

```
def op1(**kwargs):
        #body
def op2(**kwargs):
        #body

op3 = sequential(op1, op2)
```

  Where

```
def sequential(sh1, sh2, **kwargs):
    return lambda **kwargs: sh1(**kwargs) or sh2(**kwargs)
```

The **kwargs  is  a dictionary which combines the inputs  of the two  schemes, each input can be accessed through the dictionary, for example to access the value of  x, we write  kwargs[x].

**Example:**     The     Operation     Expression     transferConfirm?transferAvail; c2.printBal.Would be  equivalent  to  transferring  all  funds  from c1 to c2 and then printing the new balance of c2.

This will be mapped to using rule:

transferConfirm =  sequential (ransferAvail , c2.printBal)

### 3.8.6. Parallel composition

The parallel composition operators ‖ is used as a schema piping operator which conjoin the operation expressions by identifying and equating input variables in one operation with output variables in other operation having the same names[13]. The Parallel composition operator can be noted by

OpExp3 ::= OpExp1 ‖ OpExp2

**Rule11 :** Parallel operator in the above form will be mapped to

```
op3 = parallel(op1, op2)
  where

def  parallel(sh1, sh2, *kwargs):
    return lambda *kwargs: sh2(sh1(*kwargs))
```

Which is valid for all parallel operator, if there are variables with the same name and type exist in both first scheme output and second scheme input, first scheme call is passed to second scheme through argument which is considered a call by reference.

**Example:** The Operation Expression transferAvil?c1.withdrawAvil ‖ c2.deposit found in scheme of TwoCards class in figure2 will be mapped to:

transferAvail =c1.withdrawAvail ‖ c2.deposit

transferAvail = parallel(c1.withdrawAvail, c2.deposit)

## 4. CASE STUDY

According to the rules in section 3, the following figure shows the translation from an Object-Z specification to Python classes in figure 1 and figure 2.

```
@inv(lambda:limit in set(1000, 2000,3000)):
@inv(lambda: __instance__.balance+__instance__.limit> 0)
class CreditCard:

    limit=N
    balance=Z

    def __init__(self, b, l):
        self.balance=0

    @pre(lambda : amount <= self.balance+self.limit)
    def withdraw(self,amount):
        self.balance =  self.balance – amount

    def deposit(self,amount: N):
        self.balance =  self.balance + amount

@inv(lambda:limit in set(1000, 2000,3000)):
@inv(lambda: __instance__.balance+__instance__.limit> 0)
class TwoCreditCards:

    c1=CreditCard()
    c2=CreditCard()

    @pre(lambda :  totalbal=Z)
    def totalbal(self):
        if self.c1!=self. c2:
            totalbal=self.c1.balance+self. c2.balance

    def __init__(self):
        c1.__init__(self)
        c2.__init__(self)

    withdraw1=c1.withdraw
    transfer = conjunction(c1.withdraw, c2.deposit)
```

```
withdrawEither = choice(c1.withdraw, c2.withdraw)
transferAvail = parallel(c1.withdrawAvail, c2.deposit)

def replaceCard1(self, card :CreditCard):
    self.c1=card
```

Figure3. Translating Object-Z into Python Example

## 5. CONCLUSIONS AND FUTURE WORKS

Programming languages have different styles and paradigms with different advantages and disadvantages. Many efforts have been put into mapping C++ and java programming to Object-Z, but unfortunately the popularity and object-oriented programming paradigm takes most attention on this mapping. Python is a multi paradigms programming language, dynamic typing, high-level built-in data types, and many add-on packages, it incorporates predicate calculus, mathematical proving, set theory and a lot of libraries. In addition to that, it can be extended to contain new notations and features.

Python and Object-Z language share many similarities. Both of them are based on object-oriented paradigm, set theory and predicate calculus moreover, Python is a functional programming language which is naturally closer to formal specifications.

This work has found Python is an excellent language for developing libraries to map Object-Z specifications which is used Python's decorate capabilities to extend the language with precondition post-condition and invariant notations to check functionality constraints, In addition, writing these notations makes it possible to transform the specification into implementation with a few simple steps. In fact, these notations will to be validated through executing them instead of performing proofs. The future work could complete all other Object-Z constructs which not covered here, and could automate Object-Z formal specification transformation into implementation.